\documentclass[%
 reprint,
 superscriptaddress,
 amsmath,amssymb,
 aps,
]{revtex4-2}

\usepackage{graphicx}
\usepackage{dcolumn}
\usepackage{bm}
\usepackage{hyperref}

\begin{document}

\preprint{APS/123-QED}

\title{Eavesdropper localization for quantum and classical channels via nonlinear scattering}

\author{Alexandra Popp}
\affiliation{Max-Planck-Institute for the Science of Light, Staudtstr. 2, 91058 Erlangen, Germany}
\affiliation{Department of Physics, Friedrich-Alexander-Universit\"at Erlangen-N\"urnberg, Staudtstr. 7, 91058 Erlangen, Germany}
\affiliation{SAOT, Graduate School in Advanced Optical Technologies, Paul-Gordan-Str. 6, 91052 Erlangen, Germany}%
\author{Florian Sedlmeir}
\thanks{Currently at: Department of Physics, University of Otago, Dunedin, 3016, New Zealand}
\affiliation{Max-Planck-Institute for the Science of Light, Staudtstr. 2, 91058 Erlangen, Germany}
\affiliation{Department of Physics, Friedrich-Alexander-Universit\"at Erlangen-N\"urnberg, Staudtstr. 7, 91058 Erlangen, Germany}
\author{Birgit Stiller}
\email{birgit.stiller@mpl.mpg.de}
\affiliation{Max-Planck-Institute for the Science of Light, Staudtstr. 2, 91058 Erlangen, Germany}
\affiliation{Department of Physics, Friedrich-Alexander-Universit\"at Erlangen-N\"urnberg, Staudtstr. 7, 91058 Erlangen, Germany}
\affiliation{SAOT, Graduate School in Advanced Optical Technologies, Paul-Gordan-Str. 6, 91052 Erlangen, Germany}%
\author{Christoph Marquardt}
\email{christoph.marquardt@mpl.mpg.de}
\affiliation{Department of Physics, Friedrich-Alexander-Universit\"at Erlangen-N\"urnberg, Staudtstr. 7, 91058 Erlangen, Germany}
\affiliation{Max-Planck-Institute for the Science of Light, Staudtstr. 2, 91058 Erlangen, Germany}
\affiliation{SAOT, Graduate School in Advanced Optical Technologies, Paul-Gordan-Str. 6, 91052 Erlangen, Germany}%

\date{\today}

\begin{abstract}
Optical fiber networks are part of important critical infrastructure and known to be prone to eavesdropping attacks. Hence cryptographic methods have to be used to protect communication. Quantum key distribution (QKD), at its core, offers information theoretical security based on the laws of physics. In deployments one has to take into account practical security and resilience. The latter includes the localization of a possible eavesdropper after an anomaly has been detected by the QKD system to avoid denial-of-service. Here, we present a novel approach to eavesdropper location that can be employed in quantum as well as classical channels using stimulated Brillouin scattering. The tight localization of the acoustic wave inside the fiber channel using correlated pump and probe waves allows to discover the coordinates of a potential threat within centimeters. We demonstrate that our approach outperforms conventional OTDR in the task of localizing an evanescent outcoupling of 1$\,\%$ with cm precision inside standard optical fibers. The system is furthermore able to clearly distinguish commercially available standard SMF28 from different manufacturers, paving the way for fingerprinted fibers in high security environments.

\end{abstract}

\maketitle

                              
\section{Introduction}

Communication in optical fibers plays a crucial role in modern information society, allowing for real-time communication over large distances. However, in most cases in communication, security is of equal importance as bandwidth and distance. Therefore, for decades, researchers have devoted their time and energy to secure encryption concepts and algorithms to block eavesdroppers from their secret communication using encryption keys on messages \cite{singh1999code, shannon1948mathematical, rivest1978method}.

While established state-of-the-art encrpytion techniques rely on mathematical problems to generate and distribute encryption keys, quantum key distribution (QKD) \cite{gisin2002quantum, scarani2009security} offers a secure key exchange concept based on the laws of quantum mechanics. By monitoring the relevant parameters (quantum bit error rate in the case of discrete variable-QKD \cite{bennett1984proceedings} and excess noise in the case of continous variable-QKD \cite{ralph1999continuous}), the QKD system can bound the information gain of a possible eavesdropper and initiate appropriate counter measures with privacy amplification \cite{bennett1995generalized} to ensure security depending on the amount of channel distortion. This can also mean that key generation stops when a possible information gain is too high.

QKD thus offers the unique possibility to detect possible eavesdropping based on fundamental physics principles. However, practical security aspects and resilience have to be taken into account. In deployments one wants to avoid a denial-of-service (DoS) scenario where the key generation is rendered impossible and which has been explicitly highlighted in a comment by the NSA \cite{NSA2023}. To avoid DoS, it is necessary to localize and remove the eavesdropper from the channel after the detection of its presence. While this might be easier for a free-space QKD \cite{buttler1998practical} system, where the communication channel is often over large distances openly visible to everyone, it can be especially challenging for fiber systems \cite{fok2011optical, shaneman2004optical}, which usually come with barely accessible underground cables or many nodes which need to be kept under control and managed. This challenge exists also in purely classical communication systems.

Research has targeted this problem developing intra-fiber monitoring techniques for several decades. Today, most distributed fiber sensors are based on Rayleigh, Raman or Brillouin scattering \cite{lu2019distributed}. While all three of these offer well established sensing schemes based on the corresponding backscattering effect, Brillouin scattering, as the nonlinear effect with the highest gain coefficient in solid materials \cite{yang2020intense}, is particularly suited for high sensitivity measurements with large signal-to-noise requirements. Furthermore, due to the inelastic scatting nature, a distortion of an ambient parameter such as temperature, pressure or strain will manifest in a change of the phonon resonance frequency, which offers an additional layer of sensitivity. 

\begin{figure*}
\includegraphics{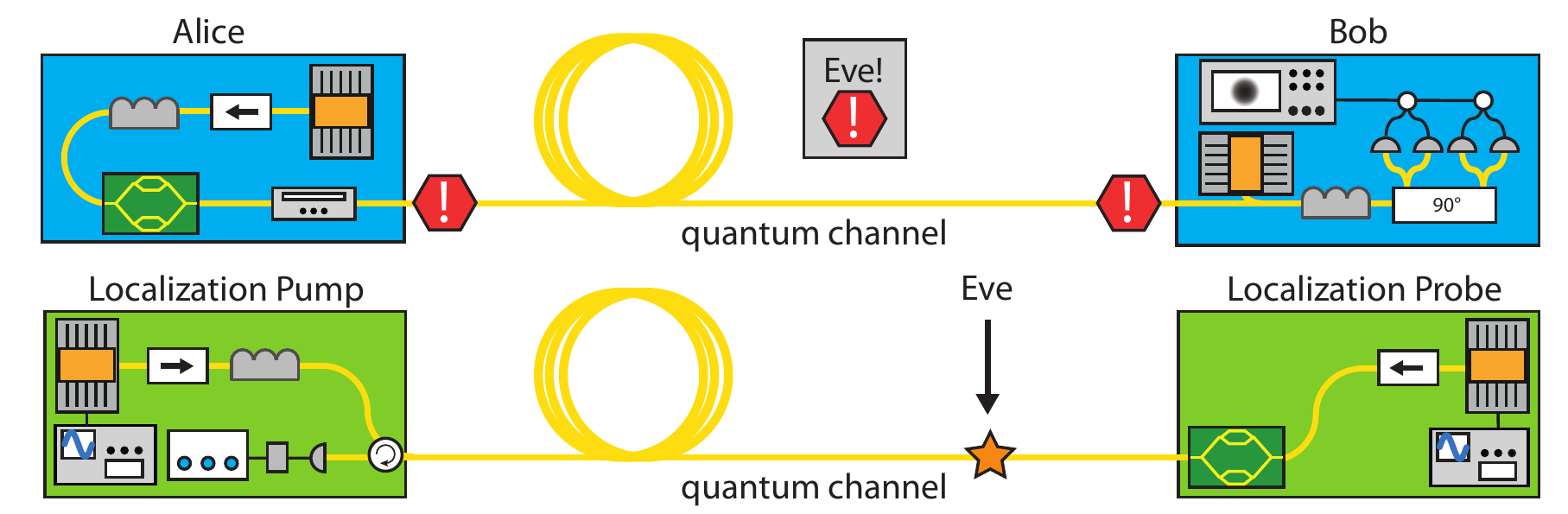}
\caption{\label{fig:Scheme} Schematic of eavesdropper localization in a quantum channel. A secret key is exchanged between the two communication parties Alice (A) and Bob (B) using a quantum key distribution (QKD) system (blue). One or both of the party members notice the presence of an eavesdropper (E) by a variation of crucial QKD parameters quantum bit error rate (QBER, DV-QKD) or excess noise (CV-QKD). If a threshold is reached, where a secure key exchange is no longer feasible, the quantum state exchange is stopped between the two parties and the channel is classified insecure. The eavesdropper localization system (green) is brought into the quantum channel. Pump and Probe are connected from A and B's access points mapping out the location of E with centimeter precision, allowing to remove E and regain the channel. QKD can be resumed.}
\end{figure*}

Most in-fiber sensing focuses on time domain methods  \cite{rao2021recent, bai2019recent}, where the sensor position is encoded in the backscattering time of short pulses. While this offers a strong advantage in sensing time and reduced complexity, this approach needs large gradients for clear distinction of features. In the case of fiber eavesdropping, this means that such systems can easily detect end facets or broken fibers \cite{fok2011optical} but reach limits when it comes to spliced connections or evanescent outcoupling \cite{iqbal2011optical}. The correlation-based approach \cite{hotate2000measurement,Zarifi2019}, which we are pursuing in this work, offers the unique advantage that it is probing the nonlinear intensity as well as spectrum with high precision and sensitivity, allowing to identify and locate an eavesdropper in a quantum or classical channel.

In this work, we present a Brillouin optical correlation domain analysis (BOCDA) system relying on cm-precision localized acoustic waves applied to a fiber channel that can be used for quantum or classical communication. We test the capability of the sensor to identify typical eavesdropping approaches on the channel. Mimicing a typical eavesdropping approach, we use evanescent outcoupling by fiber bending with different strengths between $1-10\,\%$ of transmission loss. We show that we can localize evanescent tapping with an outcoupling intensity as low as $1\,\%$ of the total channel transmission with cm precision due to its influence on the acousto-optic interaction amplitude. Additionally, we are able to localize the position of evanescent outcoupling via a commercially available in fiber tap-coupler for as low as only $1\,\%$ of coupling strengh due to its sharp influence on the acoustic resonance frequency. Furthermore, we show that standard, commercially available SMF28 single-mode fibers from three different manufacturers exhibit a distinct acousto-optic fingerprint signal. We test them when connected either as patchcord or directly spliced into the channel, allowing us to clearly identify and differentiate an attacker fiber from the channel fiber even without patchcord signature. In a final step, we realize similar measurements with a commercially available Rayleigh time domain reflectrometer (OTDR), showing that it can not reproduce our findings.

\section{Concept}
\label{chap:concept}

\begin{figure*}
\includegraphics{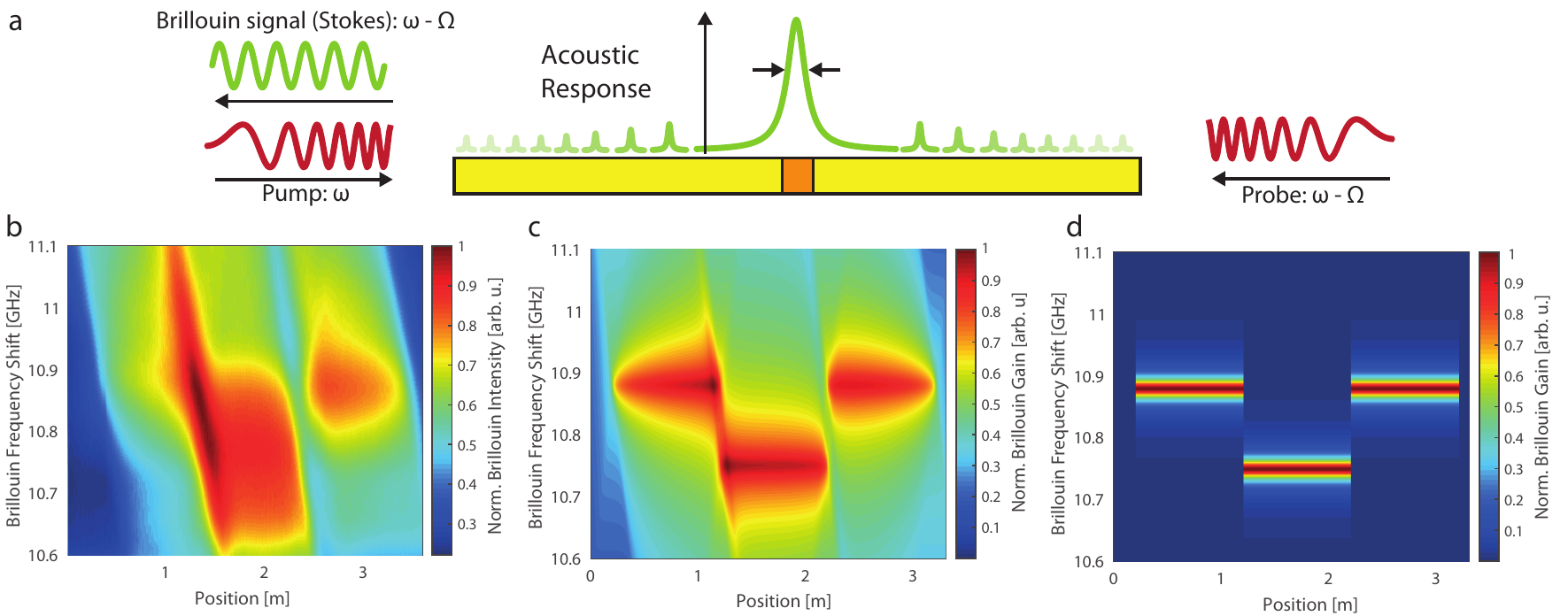}
\caption{\label{fig:BOCDA} a) Scheme of directly-frequency modulated Brillouin optical correlation domain analysis (BOCDA) in an optical waveguide. A frequency modulated pump beam (red) of optical frequency $\omega$ with modulation frequency $f_m$ is launched into the waveguide counterpropagating to an optical probe beam (red) of optical frequency $\omega - \Omega$ is modulated by the same frequency and shifted by the waveguide acoustic resonance $\Omega$. Due to frequency correlations, a stimulated acoustic response is only created at the specific correlation position $z_m(f_m)$ with a resolution $\Delta z(f_m, \Delta f)$, with the modulation bandwidth $\Delta f$. A frequency-red shifted Stokes signal is back scattered in counterpropagating direction of the pump, with environmental information encoded in its frequency and intensity. b) Typical BOCDA measurement of a 980A optical fiber (position $1.1 - 2.1\,$m embedded inside two SMF28 (positions $0.1-1.1\,$m and $2.1-3.1\,$m) fibers. Brillouin frequency shift (BFS) is shown over measurement position (arbitrary offset in m) with the normalized Brillouin intensity in color code. c) full frequency domain simulation of BOCDA measurement presented in Figure 2b d) retrieved gain spectrum without background resulting from frequency correlations.}
\end{figure*}

 When a secret key is exchanged by two parties Alice (A) and Bob (B) using QKD, it is possible that both parties detect the presence of an eavesdropper (E). When noticing this, the eavesdropper should be removed from the channel before pursuing the key exchange and thus the channel can not be used until E is localized (Figure \ref{fig:Scheme} top). We apply the well-established technique of direct frequency-modulated Brillouin optical correlation domain analysis (BOCDA) with high spatial resolution \cite{hotate2000measurement,Zarifi2019} to localize the eavesdropper.

BOCDA uses stimulated Brillouin scattering (SBS) \cite{wolff2021brillouin}, a third-order nonlinear interaction between light and acoustic phonons in a medium underlying energy and momentum conservation. The Brillouin resonance frequency ${\Omega_\mathrm{B}}$ is dependent on temperature and pressure of the host medium via the effective refractive index and acoustic velocity
\begin{align}
\label{eq:delN}
\frac{\Omega_\mathrm{B}}{2\pi} & = \frac{2}{\lambda}  \cdot n_\mathrm{eff}(p,T)  \cdot v_\mathrm{ac}(p,T).
\end{align}
In a standard configuration, the acoustic phonons are excited by a strong optical pump of frequency $\omega_\mathrm{Pump}$ together with a counter-propagating  weak optical probe detuned by the Brillouin resonance $\Omega_\mathrm{B}$ resulting in a backscattered, red-shifted Stokes interaction at $\omega_\mathrm{Stokes} = \omega_\mathrm{Pump} - \Omega_\mathrm{B}$.

In the case of directly frequency-modulated BOCDA, the counter propagating pump and probe fields are both sinusoidally frequency modulated with a modulation frequency $f_\mathrm{m}$ and bandwidth $\Delta f$ as shown in Figure \ref{fig:BOCDA}. This results in a position $z$ and time $t$ dependent beating with frequency 
\begin{align}
\label{eq:DeltaOmega}
\Delta \Omega = 2\pi \Delta f \sin(2\pi f_\mathrm{m} t) \cdot \sin(2\pi f_\mathrm{m} z/v_\mathrm{g}),
\end{align}
with the group velocity of the light in the medium $v_\mathrm{g}$. Resulting from this beating, SBS is only possible at the correlation points $z_\mathrm{m}=n v_\mathrm{g}/(2 f_\mathrm{m})$, $n \in \mathbb{N}$ with a spatial extent of
\begin{align}
\label{eq:Resolution}
\Delta z = \frac{\Delta \nu_\mathrm{B} v_\mathrm{g}}{2\pi f_\mathrm{m}\Delta f},
\end{align}
due to the necessary overlap of the probe frequency with the Brillouin resonance, where $\Delta \nu_\mathrm{B}$ is the Brillouin gain bandwidth. At all other positions, only spontaneous scattering can be observed due to the rapid change in pump-probe frequency difference according to equation \ref{eq:DeltaOmega}. Using standard single mode fibers with a Brillouin gain bandwidth of $\Delta \nu_\mathrm{B} = 27\,$MHz, a modulation bandwidth of $\Delta f = 47\,$GHz and a modulation frequency $f_\mathrm{m} = 699\,$kHz, we achieve a maximum between two different correlation points $\Delta z_\mathrm{m} = 146\,$m and a resolution of $\Delta z = 3.0\,$cm. The correlation point distance can easily be extended to several kms by decreasing the modulation frequency. The position inside the sensing target can be moved by adjusting the modulation frequency, which allows for delayline-free measurements, which is a strong benefit for communication channels which can be on the order of several kms. 

\begin{figure*}
\includegraphics{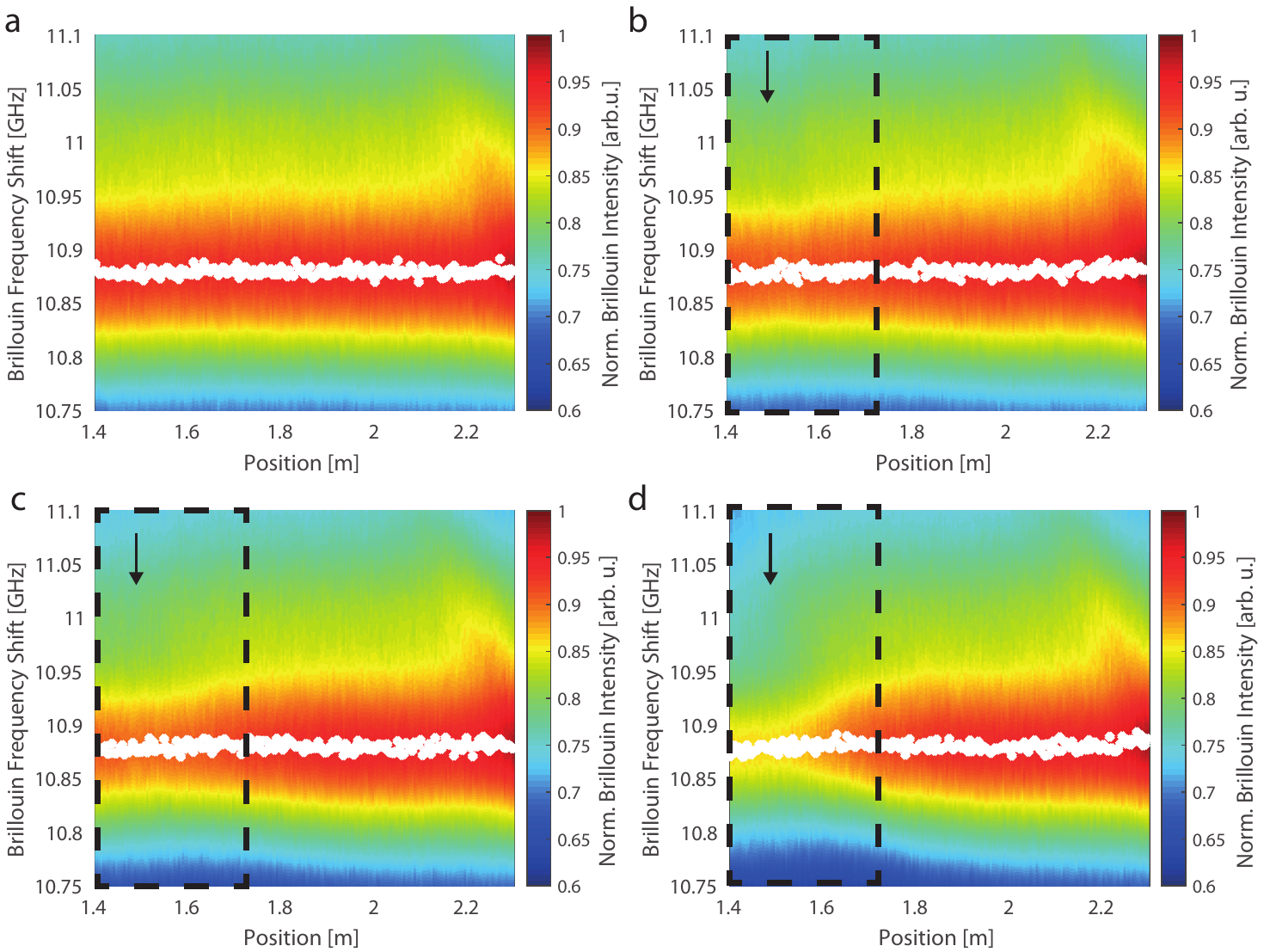}
\caption{\label{fig:bending} a) Full spectrum reference measurement of $1\,$m SMF28, BFS over position, normalized Brillouin intensity color coded. White dots denote location of maximum BFS. Steady BFS is observed with nearly constant spectral features. Spectral distortion at position $2.15\,$m results from patchcord for coupling at postion $2.35\,$m (not shown) b) Same fiber as in Figure 3a with a bending at position $1.5\,$m resulting in an evanescent coupling loss of $1\,\%$ (location marked by black box). Clear change of spectral features visible, marking the bending position. c) and d) Similar measurement as Figure 3b for a bendloss of $5\,\%$(c) and $10\,\%$(d) resulting in strongly enhanced spectral features.}
\end{figure*}

For a complete measurement as shown in Figure \ref{fig:BOCDA}b, the full Brillouin resonance is acquired for each position of the channel. For this, the probe frequency is swept over the Brillouin resonance, data acquisition is handled by a fast ADC ensuring Hz-speed measurement rates (for details and setup see supplementary material). We can employ frequency domain simulations following references \cite{yamauchi2004performance, song2018measurement} for full understanding of the spectrum (Figure \ref{fig:BOCDA}c). Furthermore, we use it for retrieval of the underlying gain spectra without the influence of the frequency correlation (Figure \ref{fig:BOCDA}d) causing the asymmetric tilt of the spectra observed in figures \ref{fig:BOCDA}b and c. Most of the time it is however sufficient to extract information directly from the raw data, especially for a channel with a smooth underlying gain profile, which is usually the case for long fibers. With this type of measurement it is now possible to gain spatially resolved information about the spectral profile of the Brillouin response as well as the intensity of all spectral components. As shown in Figure \ref{fig:BOCDA}, this allows the straight-forward distinction of fibers with slightly different refractive index. For illustration, we embed $1\,$m of commercially available 980A fiber patchord (position $1.1-2.1\,$m within two SMF28 (positions $0.1-1.1\,$m and $2.1-3.1\,$m). While both fibers support $1550\,$nm light, the 980A is clearly localized by the measurements due to the shift in Brillouin resonance by approximately $150\,$MHz. 
Another advantage of this technique is the agility of resolution and sensing distance, which can be accessed by simple variation of the frequency modulation allowing to reach resolutions from few centimeters to several 10s of meters along with a delayline-free sensing range of several kilometers \cite{wang2022recent}. 

\section{Application}

A simple, but conventionally very hard-to-detect eavesdropping approach is evanescent coupling \cite{shaneman2004optical}. Using fibers, evanescent outcoupling can easily be realized by bending the fiber either using appropriate devices or just brute-force. Already bending the fiber slightly will result in several percent of transmission loss for the channel which is accessible for E. Since this kind of bending does not produce a sharp edge, it is difficult to detect it with conventional OTDR \cite{iqbal2011optical}.

Full spectral measurements of $1\,$m SMF28 which is bent to achieve evanescent outcoupling are shown in Figure \ref{fig:bending}. A reference measurement of the undisturbed fiber is provided in Figure \ref{fig:bending}a. The measurement scheme is as explained in section \ref{chap:concept} and shown in Figure \ref{fig:BOCDA}b, the white dots in each measurement denote the location of maximum BFS. The fiber is bent around the location $1.5\,$m over a length of about $10\,$cm, inducing transmission losses of $1\,\%$ (Figure \ref{fig:bending}b), $5\,\%$ (Figure \ref{fig:bending}c) and $10\,\%$ (Figure \ref{fig:bending}d). 

\begin{figure*}
\includegraphics{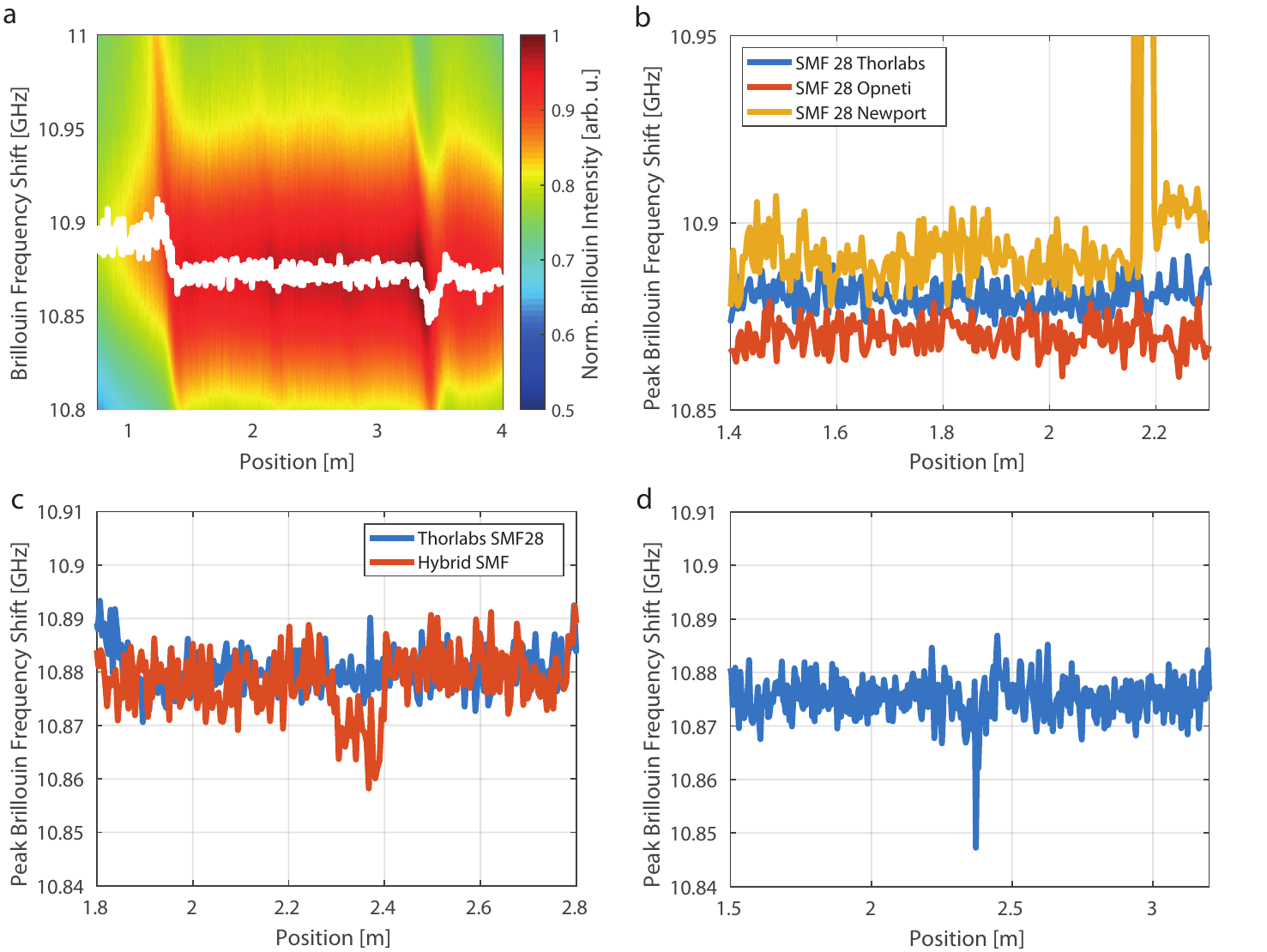}
\caption{\label{fig:Fingerprint} a) Full spectrum of Opneti SMF28 (position $1.4-3.4\,$m, BFS over position, norm. Brillouin intensity color coded, white dots denote location of maximum BFS) in between Thorlabs SMF28 (position $0.8-1.4\,$m) and Opneti SMF28 (position $3.4-4.0\,$m) all connected with FC/AP connectors. Clear localization of FC/APC connector positions (connectors C1 at $1.4\,$m and C2 at $3.4\,$m) through full spectral features as well as maximum BFS only. Clear distinction of maximum BFS frequency for fibers from different manufacturers. b) Peak BFS over position for commercially available SMF28 fibers from three manufacturers Thorlabs (blue), Opneti (red) and Newport (yellow) showing clear distinction of mean maximum BFS of several MHz. c) Comparison of maximum BFS of standard Thorlabs SMF28 patchcord (blue) with hybrid spliced patchcord (red, $6\,$cm Opneti SMF28 spliced into $1.0\,$m Thorlabs SMF28), section of Opneti fiber clearly visible by mean BFS shift. d) Maximum BFS over position of commercially available Thorlabs 99-1 in-fiber beamsplitter clearly locating the $1\,\%$ splitting position.}
\end{figure*}

For all three cases, the Brillouin intensity is locally influenced by the bendloss, showing a narrowing of the high frequency signal around the bend position due to the decreased opto-acoustic interaction amplitude. This asymmetric decrease is caused by the asymmetric shape of the BOCDA background explained in section \ref{chap:concept}, which also causes the peak around position $2.2\,$m that is visible in all measurements including the reference. As expected, the amount of intensity loss is continuously increasing proportional to bendloss. No significant change in maximum BFS is visible, confirming that the bending is not influencing the longitudinal acoustic phonons but only the opto-acoustic interaction amplitude. This means that this change is not visible by most techniques monitoring only refractive index changes, but localizable explicitly due to the large nonlinear Brillouin gain. Furthermore, it can be seen that the intensity decrease is highly localized around the bending position and, especially in the $1\,\%$-case, only a small fraction of the total intensity. Thus, the cm-resolution offers a considerable advantage not only in localization sensitivity, but also in precision.

Next, we replace the previously used SMF28 by an SMF28 patchcord from a different manufacturer. All fibers used in this case are commercially available standard SMF28 with FC/APC connection from the manufacturers Thorlabs, Opneti and Newport. We measure the Brillouin response of each of the fibers separately in between two standard SMF28 fibers, where the fiber at low position is a Thorlabs SMF28 and the fiber at high position is an Opneti SMF28 (Figure \ref{fig:Fingerprint}a). The FC/APC connectors are clearly visible in the Brillouin intensity as well as in the maximum BFS. Furthermore, we see that the maximum BFS of the Opneti SMF28 is clearly distinct from the Thorlabs SMF28. Thus, we can monitor only the maximum BFS and find significant distinction between the three manufacturers (Figure \ref{fig:Fingerprint}b) by several MHz. Therefore, we can assign a fingerprint BFS to seemingly undifferentiable commercially available fibers directly using our measurements without significant postprocessing. Furthermore, we create a hybrid SMF28 splicing $6\,$cm of Opneti SMF28 into a Thorlabs SMF28 patchcord (Figure \ref{fig:Fingerprint}c) to exclude a bias from the connection facets. Figure \ref{fig:Fingerprint}c shows that we can clearly identify the spliced piece by just monitoring the maximum BFS, which shows a difference of approximately $10\,$MHz from the rest of the hybrid fiber. Thus, we have confirmed, that even though these fibers should be nearly identical and have very similar light guiding properties, their respective refractive index or acoustic velocity (resulting from equation \ref{eq:delN}) varies enough to identify and distinguish each of them clearly.

\begin{figure*}
\includegraphics{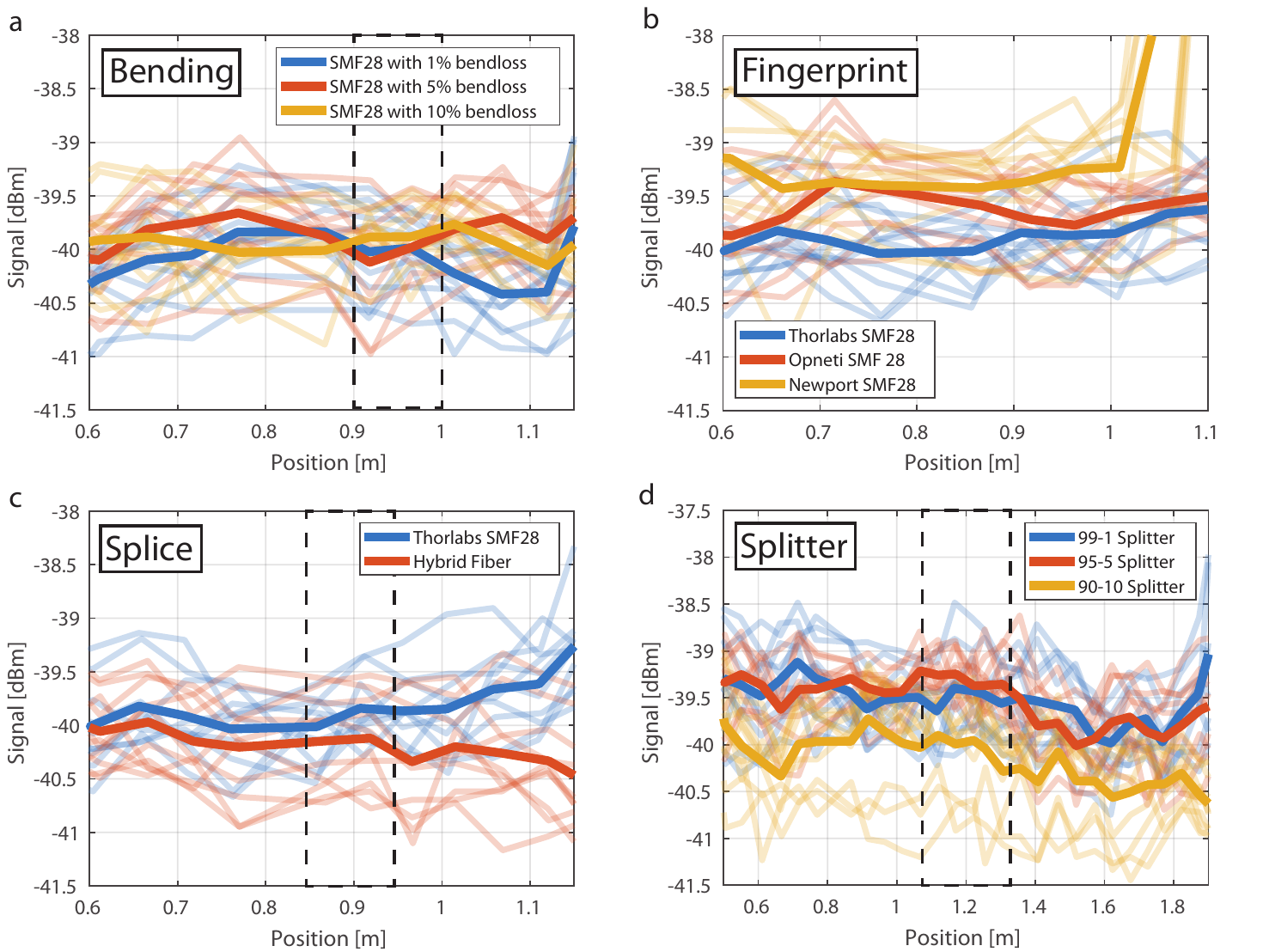}
\caption{\label{fig:OTDR} OTDR measurements using commercially available LOR-200 high resolution OTDR with $2\,$ns pulses and $2.5\,$cm sampling resolution. Saturated, solid curves of all plots mark the mean of 10 consecutive measurements, weakly saturated curves mark results of all 10 individual measurements. a) SMF28 fibers of Thorlabs (blue), Opneti (red) and Newport (yellow) as shown with BOCDA in Figure 4b. No clear distinction of fiber composition possible, significant scattering of individual measurements (weakly saturated curves). b) SMF28 with $1\,\%$ (blue), $5\,\%$ (red) and $10\,\%$ (yellow) transmission loss introduced by evanescent outcoupling similar to Figure 3. Marked area denotes bending position. No point of introduced loss is visible for any of the induced transmission losses. c) Thorlabs SMF28 (blue) and hybrid fiber with $6\,$cm Opneti SMF28 spliced in between Thorlabs SMF28 at marked position. No clear identification of spliced section, large variance of individual measurement sets (weakly saturated curves). d) In-fiber 99-1 (blue), 95-5 (red) and 90-10 (yellow) splitters. Splitting at marked position, no clear identification of splitting location or induced loss possible.}
\end{figure*}

This result has severe implications on the possibility to secure fiber networks. Showing that the difference in standard SMF28 composition is enough for clear identification of the fiber type paves the way for complete fingerprint measurements of deployed fiber channels, where parts that are added after a maintenance or offline period of the network can be clearly mapped and identified, ensuring no third party tampering with the channel. An additional layer of security could be achieved by only using fibers with a special composition, unknown to the public, when deploying new fiber networks in the future. As there are currently many future quantum fiber networks planned and deployed in Europe and around the world \cite{OpenQKD, zhang2019quantum}, our findings could be used in dedicated deployments in the near future. Furthermore, due to the high resolution of only $\Delta z= 3.00\,$cm together with the above features, our system can also be used to detect tampering inside complex fiber or waveguide based devices. This can create an additional layer of security for end-users, removing the necessity of blind trust into black boxes supplied by manufacturers.

Another tapping method is the simple optical tap-coupler \cite{shaneman2004optical}, which can split optical powers of different fractions as small as $1\,\%$ of transmission. This small fraction of tapping loss might easily be overlooked, especially in a classical system. Here, we show (Figure \ref{fig:Fingerprint}d) that it is possible to localize the interruption point as a very distinct feature in the maximum BFS.

In a final step, we compare the performance of our system to a commercially available Rayleigh OTDR. We use a LOR-200 high resolution OTDR with 2ns pulses and $2.5\,$cm sampling resolution and repeat the measurements shown in Figures \ref{fig:bending} and \ref{fig:Fingerprint}. Results are shown in Figure \ref{fig:OTDR}, where an average of 10 consecutive measurements (solid curves) as well as the results of all 10 individual measurements (weakly saturated curves) is given. The difference in absolute position is caused by the measurement method, where we define the absolute 0 of the BOCDA measurement at the start of the seed isolator, while we define the absolute zero of the OTDR measurement at the beginning of the respective test fiber. During our measurements, we do not find any signature of the loss induced by the fiber bending up to a transmission decrease of $10\,\%$, this is a significantly worse performance than considering the results in Figure \ref{fig:bending}, where a clear signature is already visible for $1\,\%$ loss. Comparing with Figure \ref{fig:Fingerprint}b, we find that first of all, no clear distinction of the individual fiber manufacturers is possible, since the scatter of the individual measurements is large. Furthermore, also the mean does not reproduce our previous findings, producing partly overlapping curves as shown in Figure \ref{fig:OTDR}b. In agreement with Figure \ref{fig:OTDR}b, it was also not possible to resolve the hybrid spliced fiber as shown in Figure \ref{fig:OTDR}c in contrast to Figure \ref{fig:Fingerprint}c. There were also no significant features caused by the splicing facets. Lastly, also no significant signature of the optical splitters was observed, while splitters with ratios up to $10\,\%$ splitting were investigated (Figure \ref{fig:OTDR}d). Note that recent research on OTDR \cite{azendorf2022distributed} has shown the sensitive detection of fiber connections via APC connectors \cite{iqbal2011optical}, however do usually not specialize in gradual changes of the refractive index.

\section{Conclusion}

In this work, we have shown a novel method for eavesdropper localization that can be applied to quantum as well as classical channels. We use localized acoustic waves created via Brillouin optical correlation domain analysis (BOCDA) to monitor standard optical fibers with cm spatial resolution, observing the complete Brillouin spectrum. This allows us to monitor the BFS caused by the longtitudinal acoustic phonons as well as the intensity of the nonlinear scattering. We clearly identify and localize an eavesdropping approach using evanescent outcoupling via fiber bending with a transmission loss down to $1\,\%$. Furthermore we show that we can clearly distinguish commercially available standard SMF28 fibers from three different manufacturers, finding that each of them has a clear fingerprint BFS distinct from the others. In addition, this can be applied inside a hybrid spliced fiber, showing that this distinction is possible over a length of $6\,$cm of spliced SMF28. Also, we are able to identify splitting by commercially available tap-couplers, clearly locating the tap-position of a $1\,\%$ splitter. In a final step, we show that it was not possible to recreate any of these findings with a commercially available OTDR. 

Acknowledging the importance of physical layer security for optical communication networks, we believe that our monitoring approach, which can be used for classical as well as quantum channels, will allow more than one additional layer of eavesdropper security for all communication systems. 

The possibility to detect evanescent eavesdropping via acoustic waves is a great advance compared to conventional OTDR as it excels in the detection of gradual changes of the refractive index as usually found in eavesdropping techniques. Furthermore, our novel finding of distinct Brillouin fingerprints for commercially available SMF28 opens up the possibility for dedicated fiber composition designs for high security application, allowing us to precisely characterize future networks and prevent fiber additions and other tampering. While the setup can be used as a stand-alone device, it can also be implemented into QKD systems, operating for example in a switching mode. This can pave the way to consistent monitoring about the installed networks in real-time. Finally, it has to be stressed, that the physical limits of the detection sensitivity and resolution are yet to be explored and can be pushed further by for example applying machine-learning techniques to detect even smaller deviations.

\begin{acknowledgments}
This research was carried out within the scope of the QuNET project, funded by the German Federal Ministry of Education and Research (BMBF) in the context of the federal government’s research framework in IT security ’Digital. Secure. Sovereign.’. We acknowledge funding from the Max Planck Society through the Independent Max Planck Research Group scheme. We thank our collegues Michael H. Frosz for providing the OTDR and K. Jaksch and A. Zarifi for fruitfull discussions. 
\end{acknowledgments}

\bibliography{bibliography}

\end{document}